\documentclass[amsmath,amssymb,a4paper]{revtex4}

\input{colordvi.tex}

\usepackage{bm}
\usepackage[mathscr]{eucal}
\def\vg{\vec{v}_{\gamma}}
\def\ve{\vec{v}_{e}}
\def\vp{\vec{v}_{p}}
\def\vb{\vec{v}_{b}}
\def\me{m_{e}}
\def\mp{m_{p}}
\def\ne{n_{e}}
\def\np{n_{p}}
\def\nb{n_{b}}
\def\dn{\delta n_{pe}}
\def\dv{\delta \vec{v}_{pe}}
\def\dvg{\delta \vec{v}_{\gamma b}}
\def\rhog{\rho_{\gamma}}
\def\Pig{\Pi_{\gamma}}

\def\deltag{\delta_{\gamma}}

\def\Dg{\Delta_{\gamma}}
\def\Db{\Delta_{b}}
\def\brhog{\bar{\rho}_{\gamma}}
\def\bnb{\bar{n}_{b}}
\def\Rbz{\bar{R}^{(0)}}

\begin{document}

\title{Electromagnetic Properties of the Early Universe}

\author{%
Keitaro Takahashi$^1$ \footnote{E-mail address:keitaro@yukawa.kyoto-u.ac.jp},  
Kiyotomo Ichiki$^2$ and 
Naoshi Sugiyama$^{3,4}$}
\affiliation{%
$^1$Yukawa Institute for Theoretical Physics, Kyoto University,
Kyoto 606-8502, Japan}
\affiliation{%
$^2$Research Center for the Early Universe, University of Tokyo,
7-3-1 Hongo, Bunkyo-ku, Tokyo 113-0033, Japan}
\affiliation{%
$^3$Department of Physics and Astrophysics, Graduate School of Science, Nagoya University,
Nagoya 464-8602, Japan}
\affiliation{%
$^4$Institute for Physics and Mathematics of the Universe, University of Tokyo, 5-1-5 Kashiwa-no-Ha, Kashiwa City,
Chiba 277-8582, Japan}

\date{\today}

\begin{abstract}
Detailed physical processes of magnetic field generation from density
fluctuations in the pre-recombination era are studied.  Solving
Maxwell equations and the generalized Ohm's law, the evolutions of the
net charge density, the electric current and the electromagnetic field
are solved.  Unlike most of previous works, we treat electrons and
photons as separate components under the assumption of tight coupling.
We find that generation of the magnetic field due to density
fluctuations takes place only from the second order of both
perturbation theory and the tight coupling approximation.

\end{abstract}

\maketitle

\section{Introduction}

Magnetic field generation from density fluctuations in the
pre-recombination era ($T \gtrsim 0.3$ eV, where $T$ is temperature
of the universe) has been investigated intensively by many authors
\cite{Hogan00,Dolgov04,Matarrese05,GopalSethi05,PRL,Science,SiegelFry06,PRD,TCA}.
Density fluctuations were generated quantum mechanically during the
inflationary epoch, evolved linearly during the radiation dominated
era, and then acted as the seed for the anisotropy of cosmic microwave
background and large scale structure of the universe.  Three important
ingredients beside dark matter and neutrinos are photons, protons and
electrons They basically behaved as a single fluid until the
recombination epoch due to strong coupling through Thomson and Coulomb
scatterings. However, because photons preferentially interact with
electrons rather than protons, there must be tiny but finite deviation
of motion between electrons and protons, that is, the net charge
density and the electric current. These are what generate magnetic fields
before recombination.

This mechanism has attracted considerable attentions because it could
give the seed fields for galactic magnetic fields. Galaxies are
observationally known to have magnetic fields of order 1 $\mu$G while
the origin has been a great mystery in modern astrophysics. It is
usually considered that if a galaxy has tiny ``seed'' magnetic fields
at its early stage, various hydrodynamical and/or plasma instabilities
would amplify the seed fields. This is known as the dynamo mechanism.
Accordingly, the problem to find out the origin of galactic magnetic
fields reduces to that of seed magnetic fields. Although there have
been many mechanisms proposed on the origin of seed fields, the
scenario considered here, which is magnetic field generation from
density fluctuations, has a great advantage compared to other
scenarios that it can give a robust evaluation of generated magnetic
fields. This is because density fluctuations themselves have already
been measured accurately and a theoretical tool to discuss them,
cosmological perturbation theory, has been established firmly. 
Note that second order density perturbations, which are the next order
of linear perturbations, are necessary to be considered for generation
of magnetic fields. It is rather lengthy and complicated to solve
second order density perturbations and this is why reliable
quantitative estimation of seed magnetic fields has not been appeared
until recent. For more information on seed magnetic fields and other
scenarios, see a comprehensive review \cite{Widrow}.

So far, the amplitudes of the generated magnetic fields estimated by
different authors are roughly consistent on the horizon scale
at recombination, although they differ by several orders on smaller scales
\cite{Hogan00,Dolgov04,Matarrese05,GopalSethi05,PRL,Science,SiegelFry06,PRD}.
Moreover, the previous studies have focused exclusively on the
magnetic field but not on other electromagnetic properties such as the
electric field, the net charge density and the electric current. In
order to understand the physical processes of magnetic field generation,
however, it is undoubtedly important to consider all of them
consistently. For example, one may (wrongly) conclude that magnetic fields
cannot be generated because tiny deviation of motion between electrons
and protons, which induce magnetic fields, could not be maintained 
due to the shorter timescale of Coulomb scattering between protons and
electrons than the one of Thomson scattering between photons and charged
particles. One will find out why this argument is not true later
in this paper. The main purpose of this paper is to clarify all physical
processes working on generation of magnetic fields via density
fluctuations and to make a physical interpretation of the results of
our previous papers \cite{PRL,Science,PRD,TCA}. We will solve Maxwell
equations and the generalized Ohm's law to express electromagnetic
quantities in terms of Thomson scattering term, which is an external
force from photons and proportional to velocity difference between photons
and charged particles.

As we mentioned earlier, photons and charged particles basically behave as
a single fluid such that density fluctuations of them evolve together.   
Because no magnetic field generation takes place in a single fluid limit,
we need some deviation from that limit. But how large deviation do we need?
To answer this question, it is helpful to consider the tight coupling
approximation \cite{PeeblesYu70}. This approximation is based on the fact
that the interaction timescale, $\tau_{\rm int}$, between two fluids is
much less than the dynamical timescale, $\tau_{\rm dyn}$. 
The coupling between two fluids is tighter for a smaller interaction
timescale and in the limit of $\tau_{\rm int}/\tau_{\rm dyn} \rightarrow 0$,
we have exact tight coupling and no velocity difference between two fluids.
Therefore it is useful to expand physical quantities with respect to
the tight coupling parameter $\tau_{\rm int}/\tau_{\rm dyn}$ in order to
estimate deviation from a single fluid.

Let us give a more specific argument. The scattering timescales for
Thomson between photons and charged particles, and Coulomb scatterings
between protons and electrons are,
\begin{eqnarray}
&& \tau_{\rm T}
   = \frac{\mp}{\sigma_{T} \rhog}
   \approx 2 \times 10^3 ~ {\rm sec} \left( \frac{1+z}{10^{5}} \right)^{-4}, \\
&& \tau_{\rm C}
   = \frac{\me}{e^2 \ne \eta}
   \approx 4 \times 10^{-3} ~ {\rm sec} \left( \frac{1+z}{10^{5}} \right)^{-3/2},
\end{eqnarray}
where $\mp$ is the proton mass, $\sigma_T$ is the Thomson cross section,
$\rhog$ is the photon energy density, $\me$ is the electron mass, 
$e$ is the electric charge, $\ne$ is the electron number density and
$z$ is the redshift. Here $\eta$ is electric resistivity as,
\begin{equation}
\eta \equiv \frac{\pi e^{2} \sqrt{m_{e}}}{T^{3/2}} \ln{\Lambda}
= 10^{-15} ~ {\rm sec} \left( \frac{1+z}{10^{5}} \right)^{-3/2}
  \left( \frac{\ln{\Lambda}}{10} \right),
\end{equation}
where $\ln{\Lambda}$ is the Coulomb logarithm.

On the other hand, the dynamical timescale of fluid may be thought as
that of acoustic oscillations.  For a given Fourier mode whose wave
number is $k$, the timescale of acoustic oscillations can be written
as $\tau_{\rm ac} = 1/kc_{\rm s}$, where $c_{\rm s}$ is the sound
velocity.  In the radiation dominated epoch, $c_{\rm s}=c/\sqrt{3}$,
where $c$ is speed of light which we take unity hereafter.
Accordingly the timescale $\tau_{\rm ac}$ can be approximately written
as $k^{-1}$.  This simply means that a Fourier mode begins to acoustically
oscillate roughly when it enters the horizon. This acoustic
oscillation remains until the fluctuation of the scale damps away
due to the finite mean free path of photons. This phenomenon, known as
Silk damping, occurs at the scale of the diffusion length of photons,
$k^{-1}_{\rm Silk}$. Deviation from tight coupling also occurs due to
the finite mean free path so that magnetic field generation will be
effective at around the Silk scale, rather than the horizon scale.  Thus,
the dynamical timescale can be thought to be roughly the Silk scale,
\begin{equation}
k^{-1}_{\rm Silk}
\approx \sqrt{\frac{\tau_{\rm cos}}{\sigma_T \ne}}
\approx 8 \times 10^{6} ~ {\rm sec} \left( \frac{1+z}{10^{5}} \right)^{-5/2},
\end{equation}
where $\tau_{\rm cos}$ is cosmological time scale, which is the inverse of
Hubble parameter, $H$,
\begin{equation}
\tau_{\rm cos} = H^{-1}
\approx 5 \times 10^{9} ~ {\rm sec} \left( \frac{1+z}{10^{5}} \right)^{-2}.
\end{equation}
Then we can have a small parameter for each scattering as,
\begin{equation}
\frac{\tau_{\rm T}}{\tau_{\rm dyn}}
= 2 \times 10^{-4} \left( \frac{1+z}{10^{5}} \right)^{-3/2}, ~~~~~
\frac{\tau_{\rm C}}{\tau_{\rm dyn}}
= 5 \times 10^{-10} \left( \frac{1+z}{10^{5}} \right),
\end{equation}
where we evaluated the dynamical timescale as $k_{\rm Silk}^{-1}$.
The former is the tight coupling parameter for Thomson scattering
and determines the magnitude of velocity difference between photons
and charged particles. On the other hand, the latter is for Coulomb
scattering and related to the magnitude of the net charge density and
the electric current. However, as we will see later, deviation of
motion between protons and electrons is suppressed more than expected
from the tight coupling approximation because protons and electrons
are coupled not only by Coulomb scattering but also by the electric
field. Further, it turns out that deviation of motion between protons
and electrons is always much smaller than that between photons and
charged particles, even though $\tau_{\rm T}$ becomes smaller than
$\tau_{\rm C}$ for $T \gtrsim 10$ keV.

Finally let us give two more important timescales. One is the inverse
of plasma frequency,
\begin{equation}
\omega_{p}^{-1} \equiv \sqrt{\frac{\me}{e^{2} \ne}}
= 2 \times 10^{-9} ~ {\rm sec} \left( \frac{1+z}{10^{5}} \right)^{-3/2},
\end{equation}
and the other is the magnetic diffusion timescale,
\begin{equation}
\tau_{\rm diff}
= \frac{\tau^2_{\rm dyn}}{\eta}
\approx 7 \times 10^{28} ~ {\rm sec} \left( \frac{1+z}{10^{5}} \right)^{-7/2}.
\end{equation}
We see the following hierarchy for various timescales:
\begin{equation}
\tau_{\rm T}, \tau_{\rm C}, \eta, \omega_p^{-1} \ll \tau_{\rm dyn}
\ll \tau_{\rm cos} \ll \tau_{\rm diff}.
\end{equation}
This hierarchy tells us that we can neglect the diffusion of the magnetic field
due to electric resistivity. Also, because we focus on the dynamics of
the scales much less than the horizon scale, general relativistic effects
are expected to be unimportant. Therefore Newtonian treatment will be
sufficient and we neglect the cosmological expansion as well.
This treatment will make our analysis rather qualitative but clear,
and is appropriate for our purpose.

This paper is organized as follows. In the next section, we derive the
generalized Ohm's law and an equation for the velocity difference
between photons and charged particles. Then in section
\ref{section:Maxwell}, combining the Ohm's law with Maxwell equations,
we solve electromagnetic quantities under the assumption of the
Thomson term regarded as an external force.  The evolution of the Thomson
term is investigated in section \ref{section:Thomson} using the tight
coupling approximation for Thomson scattering, and combining with the
result of section \ref{section:Maxwell}, electromagnetic quantities
are expressed by conventional quantities such as the photon density
fluctuation.  Finally we give discussion and summary in section
\ref{section:Discussion}.

\section{Equations of Motion \label{section:EOM}}

We start from Newtonian equations of motion for fluid densities of 
photons, protons and electrons, neglecting the cosmological expansion and
the pressure of charged particles,
\begin{eqnarray}
&& \frac{4}{3} \rhog
   \left[\partial_{t} \vg + \left( \vg \cdot \nabla \right) \vg \right]
   =  - \frac{1}{3} \left[ \nabla \rhog + \nabla \cdot (\rhog \Pig) \right]
      + \vec{C}_{\gamma p}^{\rm (T)} + \vec{C}_{\gamma e}^{\rm (T)}
      - \frac{4 \rhog}{3} \nabla \Phi,
\label{gammaEOM} \\
&& \mp \np \left[\partial_{t} \vp + \left( \vp \cdot \nabla \right) \vp \right]
   = e \np (\vec{E} +\vp \times \vec{B})
     + \vec{C}_{pe}^{\rm (C)} + \vec{C}_{p \gamma}^{\rm (T)} - \mp \np \nabla \Phi,
\label{pEOM} \\
&& \me \ne \left[\partial_{t} \ve + \left( \ve \cdot \nabla \right) \ve \right]
   = - e \ne (\vec{E} + \ve \times \vec{B})
     + \vec{C}_{ep}^{\rm (C)} + \vec{C}_{e \gamma}^{\rm (T)} - \me \ne \nabla \Phi,
\label{eEOM}
\end{eqnarray}
where $\nabla$ is a derivative with respect to spatial coordinate,
$\vec{v}_{\alpha} ~ (\alpha = \gamma, p, e)$ are fluid velocities,
$\Pig$ is anisotropic stress tensor of photons, $\np$ and $\ne$ are
proton and electron number densities, respectively, $\vec{E}$ and
$\vec{B}$ are electric and magnetic fields,respectively, and $\Phi$ is
gravitational potential. Here, $\vec{C}_{ij}^{\rm (T)}$ and
$\vec{C}_{ij}^{\rm (C)}$ are collision terms for Thomson and Coulomb
scatterings between $i$ and $j$ particles, respectively
\cite{PRL,PRD}, which are written as
\begin{eqnarray}
&& \vec{C}_{\gamma p}^{\rm (T)} = - \vec{C}_{p \gamma}^{\rm (T)}
   = - \frac{\me^{2}}{\mp^{2}} \sigma_{T} \np \rhog
       \left[ (\vg - \vp) - \frac{1}{4} \vp \cdot \Pig \right], \\
&& \vec{C}_{\gamma e}^{\rm (T)} = - \vec{C}_{e \gamma}^{\rm (T)}
   = - \sigma_{T} \ne \rhog
       \left[ (\vg - \ve) - \frac{1}{4} \ve \cdot \Pig \right], \\
&& \vec{C}_{pe}^{\rm (C)} = - \vec{C}_{ep}^{\rm (C)}
   = - e^{2} \np \ne \eta ( \vp - \ve ).
\end{eqnarray}

We will rewrite these equations of motion in terms of
center-of-mass and relative quantities of charged particles, defined as,
\begin{eqnarray}
\nb &\equiv& \frac{\np + \beta \ne}{1 + \beta}, ~~~~~
\dn \equiv \np - \ne, \label{EMquantity1} \\
\vb &\equiv& \frac{\np \vp + \beta \ne \ve}{\np + \beta \ne}, ~~~~~
\dv \equiv \vp - \ve, \label{EMquantity2}
\end{eqnarray}
and conversely,
\begin{eqnarray}
\np &=& \nb + \frac{\beta}{1+\beta} \dn, ~~~~~
\ne = \nb - \frac{1}{1+\beta} \dn, \\
\vp &=&
\vb + \left[ \frac{\beta}{1+\beta} - \frac{\beta}{(1+\beta)^2} \frac{\dn}{\nb}
      \right] \dv,
~~~~~
\ve =
\vb - \left[ \frac{1}{1+\beta} + \frac{\beta}{(1+\beta)^2} \frac{\dn}{\nb}
      \right] \dv,
\end{eqnarray}
where $\beta \equiv \me/\mp$. In terms of the new variables, 
the net electric charge
density and the electric current can be written as,
\begin{eqnarray}
\rho &=& e(\np-\ne) = e \dn, \\
\vec{j} &=& e(\np \vp - \ne \ve)
=e \left[ \nb \dv + \dn \vb - \frac{1-\beta}{1+\beta} \dn \dv
          - \frac{\beta}{(1+\beta)^2} \frac{(\dn)^2}{\nb} \dv \right].
\end{eqnarray}

Then, let us rewrite the equations of motion, Eqs. (\ref{gammaEOM}) -
(\ref{eEOM}), in terms of the center-of-mass and relative quantities,
Eqs. (\ref{EMquantity1}) and (\ref{EMquantity2}), and
$\dvg \equiv \vg - \vb$. We will keep only linear terms in $\dn$ and
$\dv$, while we will keep all nonlinear terms in $\dvg$.
The neglection of higher order terms in $\dn$ and $\dv$ will be
justified later when we solve all equations and find $\dn/\nb$ and
$\left| \dv \right|$ are much smaller than $\left| \dvg \right|$
for temperatures of interest ($\me \gtrsim T \gtrsim 0.3 ~ {\rm eV}$).
From $\me \ne \times$ Eq. (\ref{pEOM}) $- \mp \np \times$
Eq. (\ref{eEOM}), we obtain an equation for $\dv$,
\begin{eqnarray}
&& \frac{\me}{e(1+\beta)}
   \Big[ \partial_t \dv
          + \left( \vb \cdot \nabla \right) \dv
          + \left( \dv \cdot \nabla \right) \vb \Big]
\nonumber \\
&& = \vec{E} + \vb \times \vec{B} - \frac{1-\beta}{1+\beta} \dv \times \vec{B}
     - \left[ e \nb \eta + \frac{1+\beta^4}{(1+\beta)^2} \frac{\sigma_T \rhog}{e}
       \right] \dv
     - \frac{1-\beta^3}{1+\beta} \frac{\sigma_T \rhog}{e}
       \left( \dvg - \frac{1}{4} \vb \cdot \Pig \right),
\label{peEOM}
\end{eqnarray}
which can be regarded as the generalized Ohm's law.  Equation of motion
for baryon fluid is obtained from $\ne \times$ Eq. (\ref{pEOM}) $+ \np
\times$ Eq. (\ref{eEOM}):
\begin{equation}
\partial_t \vb + \left( \vb \cdot \nabla \right) \vb
= \frac{e}{(1+\beta)\mp} \dv \times \vec{B}
  + \frac{1}{1+\beta} \frac{\sigma_T \rhog}{\mp}
    \left[ (1+\beta^2) \left( \dvg - \frac{1}{4} \vb \cdot \Pig \right)
           + \frac{1-\beta^3}{1+\beta} \dv \right]
  - \nabla \Phi  .
\label{bEOM}
\end{equation}

On the other hand, Eq. (\ref{gammaEOM}) can be rewritten as,
\begin{equation}
\partial_{t} \vg + \left( \vg \cdot \nabla \right) \vg
=  - \frac{1}{4} \frac{\nabla \rhog + \nabla \cdot (\rhog \Pig)}{\rhog}
   - \frac{3}{4} \sigma_T \nb
     \left[ (1+\beta^2) \left( \dvg - \frac{1}{4} \vb \cdot \Pig \right)
             + \frac{1-\beta^3}{1+\beta} \dv \right]
   - \nabla \Phi,
\label{gammaEOM2}
\end{equation}
and from Eq. (\ref{gammaEOM2}) $-$ Eq. (\ref{bEOM}) we obtain,
\begin{eqnarray}
&& \partial_t \dvg + \left( \vg \cdot \nabla \right) \dvg
   + \left( \dvg \cdot \nabla \right) \vg
   - \left( \dvg \cdot \nabla \right) \dvg
\nonumber \\
&& = - \frac{1}{4} \frac{\nabla \rhog + \nabla \cdot (\rhog \Pig)}{\rhog}
     - \frac{e}{(1+\beta)\mp} \dv \times \vec{B}
\nonumber \\
&& ~~~
     - \frac{1+R}{1+\beta} \frac{\sigma_T \rhog}{\mp}
       \left[ (1+\beta^2) \left( \dvg - \frac{1}{4} \vb \cdot \Pig \right)
              + \frac{1-\beta^3}{1+\beta} \dv \right]
\label{gbEOM}
\end{eqnarray}
where,
\begin{equation}
R \equiv \frac{3 (\mp + \me) \nb}{4 \rhog}
\approx 4 \times 10^{-3} \left( \frac{1+z}{10^5} \right)^{-1}.
\end{equation}
Among these equations, Eqs. (\ref{peEOM}), (\ref{gammaEOM2}) and
(\ref{gbEOM}) can be chosen as independent equations which describe
the motion of the three fluids. In the context of evolution of  
cosmic microwave background (CMB) anisotropies, only Eqs. (\ref{gammaEOM2})
and (\ref{gbEOM}) without $\dv$ terms have conventionally been considered.
This is based on the assumption that protons and electrons are tightly
coupled through Coulomb interaction. As we will show later,
this assumption is valid in the sense that the $\dv$ terms
in Eqs. (\ref{gammaEOM2}) and (\ref{gbEOM}) are negligible compared
to other terms. However, it is obviously impossible to argue
electromagnetic properties of the early universe in this approach,
i.e., the complete tight coupling limit or taking only the zeroth order
of the tight coupling parameter.

In the conventional approach, all physical quantities are expanded
according to cosmological perturbation theory. At the zeroth order,
the universe is homogeneous and isotropic so that all the vector
quantities vanish. Further, because we are neglecting cosmological
expansion densities and resistivity are constant both in time and
spatial coordinates: $\rhog = \rhog^{(0)}$, $\np = \ne = \nb =
\nb^{(0)}$ and $\eta = \eta^{(0)}$. At the first order, deviations
from homogeneity and isotropy are taken into account and typical
magnitude of the deviations is about $10^{-5}$.  Density and tensor
perturbations have nonzero values which depend on positions as well as
time if they were once generated during the inflation era.
Vector-type perturbations, namely divergenceless vectors such as the
magnetic field and fluid vorticities, are absent even at the first
order since solutions of their perturbations are only decaying mode.
They can exist only if we consider the second order.  Accordingly, we
have,
\begin{eqnarray}
&& \rhog(t,\vec{x}) =
   \rhog^{(0)} + \rhog^{(1)}(t,\vec{x}) + \rhog^{(2)}(t,\vec{x}) + \cdots, \\
&& \nb(t,\vec{x}) =
   \nb^{(0)} + \nb^{(1)}(t,\vec{x}) + \nb^{(2)}(t,\vec{x}) + \cdots, \\
&& \eta(t,\vec{x}) =
   \eta^{(0)} + \eta^{(1)}(t,\vec{x}) + \eta^{(2)}(t,\vec{x}) + \cdots, \\
&& \vec{B}(t,\vec{x}) = \vec{B}^{(2)}(t,\vec{x}) + \cdots,
\end{eqnarray}
and other quantities start from the first order, although fluid
vorticities are absent at the first order, i.e., $\nabla \times
\vg^{(1)} = 0$.  In this article we will consider up to the second order
in cosmological perturbation.  We see in the above equations of motion
that the Lorentz force term, $\vb \times \vec{B}$, and the Hall term,
$\dv \times \vec{B}$, are neglected because they only appear from the
third order.

\section{Solving Maxwell + Ohm \label{section:Maxwell}}

In this section, we solve Maxwell equations and the generalized Ohm's
law to obtain electromagnetic quantities, $\rho, \vec{j}, \vec{E}$ and
$\vec{B}$. Maxwell equations and the charge conservation law are
written as,
\begin{eqnarray}
&& \nabla \cdot \vec{E} = e \dn, \\
&& \partial_{t} \vec{E} = \nabla \times \vec{B} - e (\nb \dv + \dn \vb), \\
&& \partial_{t} \vec{B} = - \nabla \times \vec{E}, \\
&& \partial_{t} \dn + \nabla \cdot (\nb \dv + \dn \vb) = 0,
\end{eqnarray}
and the generalized Ohm's law is obtained from Eq. (\ref{peEOM}) as,
\begin{equation}
\vec{E}
= \frac{\me}{e(1+\beta)}
  \left[ \partial_t \dv + \left( \vb \cdot \nabla \right) \dv
         + \left( \dv \cdot \nabla \right) \vb \right]
  + e \nb \eta_{\rm eff} \dv + \vec{C}.
\end{equation}
Here $\eta_{\rm eff}$ is an effective electric resistivity which includes
contributions from Thomson scattering,
\begin{equation}
\eta_{\rm eff}
\equiv \eta + \frac{1+\beta^4}{(1+\beta)^2} \frac{\sigma_T \rhog}{e^2 \nb}
= \eta \left[ 1 + \frac{1+\beta^4}{(1+\beta)^2} \frac{\tau_{\rm C}}{\beta \tau_{\rm T}}
        \right],
\label{resistivity}
\end{equation}
Note that the second term in Eq. (\ref{resistivity}) is dominant
for $T \gtrsim 100 ~ {\rm eV}$. On the other hand, $\vec{C}$ is
the Thomson scattering term which is regarded as an external force
in this section,
\begin{equation}
\vec{C}
\equiv \frac{1-\beta^3}{1+\beta} \frac{\sigma_T \rhog}{e}
       \left( \dvg - \frac{1}{4} \vb \cdot \Pig \right).
\end{equation}
The purpose of this section is to express $\rho, \vec{j}, \vec{E}$ and
$\vec{B}$ in terms of $\vec{C}$ up to the second order in cosmological
perturbation.  It is convenient to decompose $\vec{C}$ into scalar
and vector parts,
\begin{equation}
\vec{C} = \vec{C}_S + \vec{C}_V,
\end{equation}
where the scalar part $\vec{C}_S$ can be written by
a gradient of a function and $\vec{C}_V$ is divergenceless,
$\nabla \cdot \vec{C}_V = 0$. As we stated above, $\vec{C}_V$
is absent at the first order in cosmological perturbation.

\subsection{First order \label{subsection:Maxwell-1}}

At the first order, the magnetic field is absent and the equations are,
\begin{eqnarray}
&& \nabla \cdot \vec{E}^{(1)} = e \dn^{(1)}, \label{divE1} \\
&& \partial_{t} \vec{E}^{(1)} = - e \nb^{(0)} \dv^{(1)},
   \label{Edot1} \\
&& \partial_{t} \dn^{(1)} + \nb^{(0)} \nabla \cdot \dv^{(1)} = 0,
   \label{charge1} \\
&& \vec{E}^{(1)} = \frac{\me}{e(1+\beta)} \partial_{t} \dv^{(1)}
                   + e \nb^{(0)} \eta_{\rm eff}^{(0)} \dv^{(1)}
                   + \vec{C}^{(1)}. \label{Ohm1}
\end{eqnarray}
Let us compare the first and second terms in r.h.s. of Eq. (\ref{Ohm1}),
\begin{equation}
\frac{|\me \partial_{t} \dv^{(1)}/ e(1+\beta)|}
     {|e \nb^{(0)} \eta_{\rm eff}^{(0)} \dv^{(1)}|}
\sim \frac{\me k}{e^2 \nb^{(0)} \eta_{\rm eff}^{(0)}}
\sim \left\{
  \begin{array}{rlll}
     & k \tau_{\rm C}
     & \sim 5 \times 10^{-10}
            \left( \frac{k}{k_{\rm Silk}} \right)
            \left( \frac{1+z}{10^5} \right)
     & \hspace{1cm} (1+z \lesssim 10^6) \\
     & k \beta \tau_{\rm T}
     & \sim 10^{-10}
            \left( \frac{k}{k_{\rm Silk}} \right)
            \left( \frac{1+z}{10^7} \right)^{-3/2}
     & \hspace{1cm} (1+z \gtrsim 10^6) \\
  \end{array} \right. , \label{suppress}
\end{equation}
where we evaluated the time derivative by the wavenumber. Thus, we can neglect
the time derivative term when we consider cosmological scales at temperatures
$T \lesssim \me$:
\begin{equation}
\vec{E}^{(1)} = e \nb^{(0)} \eta_{\rm eff}^{(0)} \dv^{(1)} + \vec{C}^{(1)}.
\end{equation}
Actually, this is the leading-order tight coupling approximation for Coulomb
scattering. Taking the divergence of this equation, we have,
\begin{equation}
\eta_{\rm eff}^{(0)} \partial_t \dn^{(1)} + \dn^{(1)}
= \frac{1}{e} \nabla \cdot \vec{C}^{(1)}.
\end{equation}
We can neglect the first term of l.h.s. because
\begin{equation}
\frac{\left| \eta_{\rm eff}^{(0)} \partial_t \dn^{(1)} \right|}{|\dn^{(1)}|}
\sim k \eta_{\rm eff}^{(0)}
\sim \left\{
  \begin{array}{rlll}
     & k \eta
     & \sim 10^{-22}
            \left( \frac{k}{k_{\rm Silk}} \right)
            \left( \frac{1+z}{10^5} \right)
     & \hspace{1cm} (1+z \lesssim 10^6) \\
     & k \eta \frac{\tau_{\rm C}}{\beta \tau_{\rm T}}
     & \sim 4 \times 10^{-18}
            \left( \frac{k}{k_{\rm Silk}} \right)
            \left( \frac{1+z}{10^7} \right)^{7/2}
     & \hspace{1cm} (1+z \gtrsim 10^6) \\
  \end{array} \right. ,
\label{suppress2}
\end{equation}
thus we have,
\begin{equation}
\dn^{(1)} \approx \frac{1}{e} \nabla \cdot \vec{C}^{(1)}.
\end{equation}
Substituting this into Eq. (\ref{charge1}), we can solve for the velocity difference,
\begin{equation}
\dv^{(1)} = - \frac{1}{e \nb} \partial_t \vec{C}^{(1)}.
\end{equation}
In general, the rotational part of $\dv^{(1)}$ cannot be determined from
Eq. (\ref{charge1}), but it should vanish at the first order.
Then, we obtain the electric field from the Ohm's law as,
\begin{equation}
\vec{E}^{(1)} = \vec{C}^{(1)},
\end{equation}
where we neglected $\eta_{\rm eff}^{(0)} \partial_t \vec{C}^{(1)}$ term.
Thus, at the first order, we have,
\begin{eqnarray}
&& \dn^{(1)} = \frac{1}{e} \nabla \cdot \vec{C}^{(1)}, \\
&& \dv^{(1)} = - \frac{1}{e \nb^{(0)}} \partial_t \vec{C}^{(1)}, \\
&& \vec{E}^{(1)} = \vec{C}^{(1)}, \label{E1}
\end{eqnarray}
and we see they also satisfy Eq. (\ref{Edot1}).  Correspondingly, the
net charge density and the electric current can be expressed as,
\begin{eqnarray}
&& \rho^{(1)} = \nabla \cdot \vec{C}^{(1)} \\
&& \vec{j}^{(1)} = - \partial_t \vec{C}^{(1)}.
\end{eqnarray}

Before we proceed to the second order, let us explain the meaning of
the approximations in Eqs. (\ref{suppress}) and (\ref{suppress2}). If
we take the divergence of the Ohm's law (\ref{Ohm1}), without neglecting
the time derivative term, we have,
\begin{equation}
\frac{1}{(1+\beta) \omega_p^2} \partial_t^2 \dn^{(1)}
+ \eta_{\rm eff}^{(0)} \partial_t \dn^{(1)}
+ \dn^{(1)}
= \frac{1}{e} \nabla \cdot \vec{C}^{(1)}.
\label{rhoEOM}
\end{equation}
Eq. (\ref{rhoEOM}) describes the dynamics of charge separation,
$\dn^{(1)}$, and can be seen as an equation of damped oscillation
with an external force. There are two key timescales. One is the
timescale of oscillation, $\omega_p^{-1}$, and another is the damping
timescale, $1/(\omega_p^2 \eta_{\rm eff}) \sim \tau_{\rm C}$.
Even though the charge separation and its time derivative are
zero initially (outside the horizon), the source term induces
the oscillation whose timescale is $\omega_p^{-1}$.
Then the oscillation is damped within the timescale $\tau_{\rm C}$,
and charge separation relaxes into the equilibrium value which is
nonzero due to the presence of the source term,
$e \dn^{(1)} = \nabla \cdot \vec{C}^{(1)}$. Eq. (\ref{E1}) tells us
that the force from photons balances with electric field in this state.
Because we are focusing on the dynamics of cosmological timescale,
it is enough to consider the equilibrium state. The first approximation,
Eq. (\ref{suppress}), corresponds to the neglection of the first
term in l.h.s. of Eq. (\ref{rhoEOM}), that is, the neglection of
plasma oscillation. Likewise, the second approximation,
Eq. (\ref{suppress2}), corresponds to the neglection of the
second term in l.h.s. of Eq. (\ref{rhoEOM}). Note that this
neglection leads to the absence of magnetic diffusion, that is,
magnetic diffusion is important only for very small scales. Thus,
both approximations are valid in our context.  Similar approximations
will also be used at the second order below.

\subsection{Second order}

At the second order, the equations are,
\begin{eqnarray}
&& \nabla \cdot \vec{E}^{(2)} = e \dn^{(2)}, \label{divE2} \\
&& \partial_{t} \vec{E}^{(2)}
   = \nabla \times \vec{B}^{(2)}
     - e \left( \nb^{(0)} \dv^{(2)} + \nb^{(1)} \dv^{(1)} + \dn^{(1)} \vb^{(1)} \right),
\label{rotB2} \\
&& \partial_{t} \vec{B}^{(2)}
   = - \nabla \times \vec{E}^{(2)}, \label{Bdot2} \\
&& \partial_{t} \dn^{(2)}
   + \nabla \cdot \left( \nb^{(0)} \dv^{(2)} + \nb^{(1)} \dv^{(1)} + \dn^{(1)} \vb^{(1)} \right)
   = 0, \label{charge2} \\
&& \vec{E}^{(2)}
   = \frac{\me}{e(1+\beta)}
     \left[ \partial_t \dv^{(2)} + \left( \vb^{(1)} \cdot \nabla \right) \dv^{(1)}
            + \left( \dv^{(1)} \cdot \nabla \right) \vb^{(1)} \right] \nonumber \\
&& ~~~~~~~~~~
     + e \nb^{(0)} \eta_{\rm eff}^{(0)}
       \left( \dv^{(2)} + \frac{\nb^{(1)}}{\nb^{(0)}} \dv^{(1)}
              + \frac{\eta_{\rm eff}^{(1)}}{\eta_{\rm eff}^{(0)}} \dv^{(1)} \right)
     + \vec{C}^{(2)}.
   \label{Ohm2}
\end{eqnarray}
As in the case of the first order, the terms in the bracket $[\cdots]$ of Eq. (\ref{Ohm2})
are suppressed by the factor in Eq. (\ref{suppress}) and can be neglected. Further,
the last two terms in the parenthesis $(\cdots)$ of Eq. (\ref{Ohm2}) can be neglected
compared to $C^{(2)}$ because,
\begin{equation}
\left| e \nb^{(0)} \eta_{\rm eff}^{(0)} \frac{\nb^{(1)}}{\nb^{(0)}} \dv^{(1)} \right| 
\sim \left| e \nb^{(0)} \eta_{\rm eff}^{(0)}
            \frac{\eta_{\rm eff}^{(1)}}{\eta_{\rm eff}^{(0)}} \dv^{(1)} \right|
\sim \left| k \eta_{\rm eff}^{(0)} \vec{C}^{(2)} \right|
\ll \left| \vec{C}^{(2)} \right|,
\end{equation}
thus we have simplified second-order Ohm's law as,
\begin{equation}
\vec{E}^{(2)} = e \nb^{(0)} \eta_{\rm eff}^{(0)} \dv^{(2)} + \vec{C}^{(2)}.
\end{equation}
As we did at the first order, we take the divergence of this equation,
\begin{equation}
\eta_{\rm eff}^{(0)} \partial_t \dn^{(2)} + \dn^{(2)}
= \frac{1}{e} \nabla \cdot \vec{C}^{(2)},
\end{equation}
and neglecting the first term, we obtain,
\begin{equation}
\dn^{(2)} \approx \frac{1}{e} \nabla \cdot \vec{C}^{(2)}.
\end{equation}
Then we can solve for $\dv^{(2)}$ from Eq. (\ref{charge2}) as,
\begin{equation}
\dv^{(2)} = - \frac{1}{e \nb^{(0)}}
              \left[ \partial_t
                     \left( \vec{C}^{(2)}_S - \frac{\nb^{(1)}}{\nb^{(0)}} \vec{C}^{(1)} \right)
                     + (\nabla \cdot \vec{C}^{(1)}) \vb^{(1)} \right]
            + \nabla \times \vec{D}^{(2)},
\end{equation}
where $\vec{D}^{(2)}$ is an undetermined vector. From Eq. (\ref{Ohm2}), we have,
\begin{equation}
\vec{E}^{(2)} = \vec{C}^{(2)}
                + e \nb^{(0)} \eta^{(0)} \nabla \times \vec{D}^{(2)}.
\end{equation}
Here we neglected terms of order $k \eta_{\rm eff}^{(0)}$.
The magnetic field can be obtained from Eq. (\ref{Bdot2}),
\begin{eqnarray}
\vec{B}^{(2)} &=& - \int dt ~ \nabla \times \vec{E}^{(2)} \\
&=& - \int dt ~
      \left[ \nabla \times \vec{C}^{(2)}
             + e \nb^{(0)} \eta^{(0)} \nabla \times \nabla \times \vec{D}^{(2)}
      \right],
\end{eqnarray}
and the vector $\vec{D}^{(2)}$ is determined by Eq. (\ref{rotB2}):
\begin{eqnarray}
&& e \nb^{(0)}
   \left[ \eta_{\rm eff}^{(0)} \partial_t + 1 \right]
   \nabla \times \vec{D}^{(2)}
   + \int dt ~ e \nb^{(0)} \eta_{\rm eff}^{(0)}
               \nabla \times \nabla \times \nabla \times \vec{D}^{(2)}
\nonumber \\
&& = - \int dt ~ \nabla \times \nabla \times \vec{C}^{(2)}
     - \partial_t \vec{C}^{(2)}_V.
\label{EOMD}
\end{eqnarray}
Evaluating time and spatial derivatives as wavenumber $k$ and time integration as $1/k$,
we pick up only dominant terms:
\begin{equation}
\nabla \times \vec{D}^{(2)}
= - \frac{1}{e \nb^{(0)}} \int dt ~ \nabla \times \nabla \times \vec{C}^{(2)}
  - \frac{1}{e \nb^{(0)}} \partial_t \vec{C}^{(2)}_V.
\label{solD}
\end{equation}
Thus we obtained all the second-order quantities in terms of $\vec{C}^{(2)}$,
\begin{eqnarray}
&& \dn^{(2)} = \frac{1}{e} \nabla \cdot \vec{C}^{(2)}, \\
&& \dv^{(2)} = - \frac{1}{e \nb^{(0)}}
                 \left[ \partial_t
                        \left( \vec{C}^{(2)} - \frac{\nb^{(1)}}{\nb^{(0)}} \vec{C}^{(1)} \right)
                        + (\nabla \cdot \vec{C}^{(1)}) \vb^{(1)}
                        + \int dt ~ \nabla \times \nabla \times \vec{C}^{(2)}
                 \right], \\
&& \vec{E}^{(2)} = \vec{C}^{(2)}, \\
&& \vec{B}^{(2)} = - \int dt ~ \nabla \times \vec{C}^{(2)},
\end{eqnarray}
and correspondingly,
\begin{eqnarray}
&& \rho^{(2)} = \nabla \cdot \vec{C}^{(2)}, \\
&& \vec{j}^{(2)} = - \partial_t \vec{C}^{(2)}
                   - \int dt ~ \nabla \times \nabla \times \vec{C}^{(2)}.
\end{eqnarray}
Combining the first and second order results, we have,
\begin{eqnarray}
&& \rho = \nabla \cdot \vec{C}, \\
&& \vec{j} = - \partial_t \vec{C}
             - \int dt ~ \nabla \times \nabla \times \vec{C},
   \label{solj} \\
&& \vec{E} = \vec{C}, \\
&& \vec{B} = - \int dt ~ \nabla \times \vec{C},
\end{eqnarray}
up to the second order. This is one of our main results. Here it
should be noted that the vector-type perturbation is absent at the first
order so that $\nabla \times \vec{C}^{(1)} = 0$. 

Now we expressed the electromagnetic quantities in terms of
the Thomson term $\vec{C}$.  We see that the electric current has
two contributions in Eq. (\ref{solj}) and the first and second terms
balance with the displacement current and rotation of the magnetic field
in (\ref{rotB2}), respectively.
Since the Thomson term $\vec{C}$ is the velocity difference between
photon and baryon fluctuations, it suffers from Silk damping when
$k^{-1}_{\rm Silk}$ exceeds $k^{-1}$, that is, $\vec{C} \rightarrow 0$
for $k^{-1} < k^{-1}_{\rm Silk}$. The above solution tells us that the
electric current and the magnetic field keep non zero values even for
$\vec{C} = 0$ while the net charge density and the electric field vanish. This
is because the diffusion timescale of the magnetic field is so much larger
than the dynamical timescale that the magnetic field, once generated, would
not damp away and be maintained by the residual electric current (the second
term in Eq. (\ref{solj})) even after the source term would disappear.

Before concluding this section let us comment on the earlier work of
\cite{SiegelFry06}. The difference in the treatment between theirs and
ours is that we have included the displacement current in 
Eq. (\ref{rotB2}) while they have omitted it. We found above that the
displacement current leads to the second term in (\ref{solj}) and it is
what maintains the magnetic fields after their generation.
We attribute the difference in the resultant magnetic fields at small
scales between \cite{SiegelFry06} and ours to this contribution from the
displacement current. 


\section{Evolution of Thomson term \label{section:Thomson}}

In this section, we follow the evolution of the Thomson term to
discuss magnetic field generation in the language of the tight
coupling approximation.  This approach was first studied in \cite{TCA}
and it was shown that magnetic field generation is absent at the first
order in the tight coupling approximation and starts from the second
order. We will confirm this fact and, furthermore, solve the equations
explicitly to express the electromagnetic quantities in terms of the
conventional quantities such as the density fluctuations of
photons. For simplicity, we will ignore the anisotropic stress of
photons hereafter. This treatment makes our analysis rather qualitative but it
would be helpful in order to see the dynamics of the tight coupling
approximation clearly and interpret the numerical analysis of our
previous papers \cite{Science,PRD}.

When we consider magnetic field generation, a particularly important
quantity is the rotation of the Thomson term,
\begin{equation}
\nabla \times \vec{C}^{(2)}
= \frac{1-\beta^3}{1+\beta} \frac{\sigma_T \rhog^{(0)}}{e}
  \left[ \frac{\nabla \rhog^{(1)}}{\rhog^{(0)}} \times \dvg^{(1)}
         + \nabla \times \dvg^{(2)} \right].
\label{slip-vorticity}
\end{equation}
Here we ignored the anisotropic stress of photons. Thus, we need to
solve for $\dvg$ up to the second order in cosmological perturbation.
In Eq. (\ref{slip-vorticity}) the first term of the r.h.s. is the
vector product of the gradient of the photon energy density and the
velocity difference between photons and baryons. This was called the "slip
term" in \cite{Science,PRD} and evaluated numerically. On the other
hand, the second term is the vorticity difference between photons and
baryons and has not been evaluated so far.  Although it is hard to
compute this term in the framework of general relativistic
perturbation theory, we can evaluate it in our simplified formalism.

The evolution equation of $\dvg$, Eq. (\ref{gbEOM}), without the
anisotropic stress of photons is,
\begin{equation}
\partial_t \dvg + \left( \vg \cdot \nabla \right) \dvg
+ \left( \dvg \cdot \nabla \right) \vg
- \left( \dvg \cdot \nabla \right) \dvg
= - \frac{1}{4} \frac{\nabla \rhog}{\rhog}
  - \frac{1+R}{1+\beta} \frac{\sigma_T \rhog}{\mp}
    \left[ (1+\beta^2) \dvg + \frac{1-\beta^3}{1+\beta} \dv \right].
\end{equation}
We can compare the relative importance of the collision terms from the result
of the previous section,
\begin{equation}
\frac{\left| \dv \right|}{\left| \dvg \right|}
\sim \frac{\sigma_T \rhog k}{e^2 \nb}
= \frac{k}{\omega_p^2 \tau_{\rm T}}
\sim 5 \times 10^{-28} \left( \frac{k}{k_{\rm Silk}} \right)
                         \left( \frac{1+z}{10^5} \right)^{7/2}.
\end{equation}
Therefore, we can neglect the $\dv$ term and we have,
\begin{equation}
\partial_t \dvg + \left( \vec{v} \cdot \nabla \right) \dvg
+ \left( \dvg \cdot \nabla \right) \vec{v}
- \left( \dvg \cdot \nabla \right) \dvg
= - \frac{1}{4} \frac{\nabla \rhog}{\rhog}
  - \frac{1+\beta^2}{1+\beta} (1+R) \frac{\sigma_T \rhog}{\mp} \dvg.
\label{gbEOM2}
\end{equation}
We have also confirmed the conventional assumption in the context of
evolution of CMB anisotropies that protons and electrons are so
tightly coupled that we can treat them as a single fluid as long as we
consider the dynamics of photons and baryons.

We will solve Eq. (\ref{gbEOM2}) by employing the tight coupling
approximation \cite{PeeblesYu70,TCA}. As we mentioned in the
introduction, this approximation makes use of the fact that the
scattering timescale, $\tau_{\rm T}$, is much shorter than the
dynamical timescale, $k^{-1}$, so that the deviation of the motion of
the two fluids is very small. The expansion parameter is the ratio of
two timescales,
\begin{equation}
k \tau_{\rm T}
= 2 \times 10^{-4} \left( \frac{k}{k_{\rm Silk}} \right)
                   \left( \frac{1+z}{10^{5}} \right)^{-3/2}.
\label{TCAparameter}
\end{equation}
Here we define the deviation of photon and baryon distributions from adiabatic
distribution by,
\begin{equation}
\rhog = \brhog (1 + \Dg), ~~~
\nb = \bnb (1 + \Db),
\label{def-Delta}
\end{equation}
respectively. These two quantities, $\Dg$ and $\Db$, are assumed to be small
and we expand them by the tight coupling parameter, Eq. (\ref{TCAparameter}),
as,
\begin{equation}
\Dg = \Dg^{(I)} + \Dg^{(II)} + \cdots, ~~~~~
\Db = \Db^{(I)} + \Db^{(II)} + \cdots.
\label{expansion-d}
\end{equation}
Fluid velocities are also expanded as,
\begin{equation}
\vg = \vec{v} + \vg^{(I)} + \vg^{(II)} + \cdots, ~~~~~
\vb = \vec{v} + \vb^{(I)} + \vb^{(II)} + \cdots, ~~~~~
\dvg = \dvg^{(I)} + \dvg^{(II)} + \cdots,
\label{expansion-v}
\end{equation}
where $\vec{v}$ is the common velocity of photons and baryons at the zeroth
order and $\dvg^{(i)} = \vg^{(i)} - \vb^{(i)}, (i = I, II, \cdots)$.
It should be noted that this expansion is independent of cosmological
perturbation and each term in Eqs. (\ref{expansion-d}) and (\ref{expansion-v})
can be expanded with respect to cosmological perturbation, e.g.
$\Dg^{(I)} = \Dg^{(I,1)} + \Dg^{(I,2)} + \cdots$.

Our purpose here is to express $\dvg$ and then the electromagnetic quantities
explicitly in terms of the conventional quantities such as $\brhog^{(1)}$.

\subsection{Continuity Equations}

To solve the equation of motion (\ref{gbEOM2}), we need to consider
the continuity equations of photons and baryons,
\begin{eqnarray}
&& \partial_t \rhog + \left( \vg \cdot \nabla \right) \rhog
   + \frac{4 \rhog}{3} \nabla \cdot \vg
   = 0, \\
&& \partial_t \nb + \left( \vb \cdot \nabla \right) \nb + \nb \nabla \cdot \vb
   = 0.
\end{eqnarray}
At the zeroth order of the tight coupling approximation, these
equations reduce to,
\begin{eqnarray}
&& \partial_t \brhog + \left( \vec{v} \cdot \nabla \right) \brhog
   + \frac{4 \brhog}{3} \nabla \cdot \vec{v}
   = 0, \\
&& \partial_t \bar{n}_b + \left( \vec{v} \cdot \nabla \right) \bar{n}_b
   + \bar{n}_b \nabla \cdot \vec{v}
   = 0,
\end{eqnarray}
which are combined to obtain,
\begin{equation}
\left( \partial_t + \vec{v} \cdot \nabla \right)
\left( \frac{\bar{n}_b}{\brhog^{3/4}} \right) = 0.
\label{TCA0density}
\end{equation}
This shows that fluctuations of photons and baryons behave adiabatically
at this order, as we expected.

As we will see later, we need the relation between $\Dg$ and $\Db$
at the first order both in the tight coupling approximation and cosmological
perturbation to discuss magnetic field generation. At this order,
the continuity equations for photons and baryons are,
\begin{equation}
\partial_t \Dg^{(I,1)} + \frac{4}{3} \nabla \cdot \vg^{(I,1)} = 0, ~~~~~
\partial_t \Db^{(I,1)} + \nabla \cdot \vb^{(I,1)} = 0,
\end{equation}
respectively, and then we have,
\begin{equation}
\Db^{(I,1)} = \frac{3}{4} \Dg^{(I,1)} + \int dt ~ \nabla \cdot \dvg^{(I,1)}.
\label{TCA1density}
\end{equation}

\subsection{Equation of Motion}

Substituting the expansion Eq. (\ref{expansion-d}) into Eq. (\ref{gbEOM2}),
we have,
\begin{equation}
\partial_t \dvg + \left( \vg \cdot \nabla \right) \dvg
+ \left( \dvg \cdot \nabla \right) \vg
- \left( \dvg \cdot \nabla \right) \dvg
= - \frac{1}{4} \left( \frac{\nabla \brhog}{\brhog} + \nabla \Dg \right)
  - \nu \dvg,
\end{equation}
where $\nu$ is the collision frequency,
\begin{equation}
\nu \equiv \frac{1+\beta^2}{1+\beta} (1+R) \frac{\sigma_T \rhog}{\mp}
= \bar{\nu} (1 + \Dg)
  \left[ \frac{1}{1+\bar{R}} + \frac{\bar{R}}{1+\bar{R}} \frac{1+\Db}{1+\Dg} \right],
\end{equation}
and barred quantities are the zeroth order of 
the tight coupling approximation:
\begin{equation}
\bar{\nu} \equiv \frac{1+\beta^2}{1+\beta} (1+\bar{R}) \frac{\sigma_T \brhog}{\mp},
~~~~~
\bar{R} \equiv \frac{3 (\mp + \me) \bnb}{4 \brhog}.
\end{equation}

At the zeroth order in tight coupling approximation, the equation of motion
reduces to
\begin{equation}
0 = - \frac{1}{4} \frac{\nabla \brhog}{\brhog} - \bar{\nu} \dvg^{(I)} .
\end{equation}
This equation can be solved up to the second order in cosmological perturbation:
\begin{eqnarray}
&& \dvg^{(I,1)}
   = - \frac{1}{4 \bar{\nu}^{(0)}} \frac{\nabla \brhog^{(1)}}{\brhog^{(0)}},
\label{TCAsol11} \\
&& \dvg^{(I,2)}
   = - \frac{1}{4 \bar{\nu}^{(0)}}
       \left[ \frac{\nabla \brhog^{(2)}}{\brhog^{(0)}}
              - \left( \frac{\brhog^{(1)}}{\brhog^{(0)}}
                       + \frac{\bar{\nu}^{(1)}}{\bar{\nu}^{(0)}}
                \right) \frac{\nabla \brhog^{(1)}}{\brhog^{(0)}}
       \right].
\label{TCAsol12}
\end{eqnarray}
We can show that the rotation of the Thomson term
(\ref{slip-vorticity}), which is the source term of the magnetic
field, vanishes at this order, denoting the following relation
obtained from the adiabaticity condition, Eq. (\ref{TCA0density}),
\begin{equation}
\frac{\bar{\nu}^{(1)}}{\bar{\nu}^{(0)}}
= \frac{4 + 3 \bar{R}^{(0)}}{4(1 + \bar{R}^{(0)})}
  \frac{\brhog^{(1)}}{\brhog^{(0)}}.
\label{nu0}
\end{equation}

Let us now consider the next order of the tight coupling approximation 
in order to argue magnetic field generation.
The first order equation is,
\begin{equation}
\partial_t \dvg^{(I)} + \left( \vec{v} \cdot \nabla \right) \dvg^{(I)}
+ \left( \dvg^{(I)} \cdot \nabla \right) \vec{v}
= - \frac{1}{4} \nabla \Dg^{(I)} - \bar{\nu} \dvg^{(II)} - \nu^{(I)} \dvg^{(I)},
\end{equation}
and this can be solved as,
\begin{equation}
\dvg^{(II,1)} = - \frac{1}{4 \bar{\nu}^{(0)}} \nabla \Dg^{(I,1)}
                + \frac{1}{4 (\bar{\nu}^{(0)})^2}
                  \frac{\partial_t \nabla \brhog^{(1)}}{\brhog^{(0)}},
\end{equation}
\begin{eqnarray}
\dvg^{(II,2)} &=&
- \frac{1}{4 \bar{\nu}^{(0)}}
  \left[ \nabla \Dg^{(I,2)} - \frac{\bar{\nu}^{(1)}}{\bar{\nu}^{(0)}} \Dg^{(I,1)}
         - \frac{\nu^{(I,1)}}{\bar{\nu}^{(0)}}
           \frac{\nabla \brhog^{(1)}}{\brhog^{(0)}}
  \right]
\nonumber \\
&&
+ \frac{1}{4 (\bar{\nu}^{(0)})^2}
  \left[ \frac{\partial_t \nabla \brhog^{(2)}}{\brhog^{(0)}}
         - \left( \frac{\brhog^{(1)}}{\brhog^{(0)}}
                  + \frac{2 \bar{\nu}^{(1)}}{\bar{\nu}^{(0)}}
           \right) \frac{\partial_t \nabla \brhog^{(1)}}{\brhog^{(0)}}
         - \left( \frac{\partial_t \brhog^{(1)}}{\brhog^{(0)}}
                  + \frac{\partial_t \bar{\nu}^{(1)}}{\bar{\nu}^{(0)}} \right)
           \frac{\nabla \brhog^{(1)}}{\brhog^{(0)}}
  \right.
\nonumber \\
&& ~~~~~~~~~~~~~~~~~ \left.
         + (\vec{v} \cdot \nabla) \frac{\nabla \brhog^{(1)}}{\brhog^{(0)}}
         + \left( \frac{\nabla \brhog^{(1)}}{\brhog^{(0)}} \cdot \nabla \right)
           \vec{v}
   \right],
\end{eqnarray}
where we substituted Eqs. (\ref{TCAsol11}) and (\ref{TCAsol12}).
We see that we have non-zero contribution for the slip term and vorticity
difference at this order as 
\begin{eqnarray}
\frac{\nabla \rhog^{(1)}}{\rhog^{(0)}} \times \dvg^{(1)}
&=& \frac{\nabla \brhog^{(1)}}{\brhog^{(0)}} \times \dvg^{(II,1)}
    + \nabla \Dg^{(I,1)} \times \dvg^{(I,1)} \nonumber \\
&=& \frac{1}{4 (\bar{\nu}^{(0)})^2}
    \frac{\nabla \brhog^{(1)}}{\brhog^{(0)}} \times
    \frac{\partial_t \nabla \brhog^{(1)}}{\brhog^{(0)}},
\end{eqnarray}
\begin{eqnarray}
\nabla \times \dvg^{(2)}
&=& \frac{1}{4 \bar{\nu}^{(0)}}
    \left[ \frac{\nabla \bar{\nu}^{(1)}}{\bar{\nu}^{(0)}} \times \Dg^{(I,1)}
           + \frac{\nabla \nu^{(I,1)}}{\bar{\nu}^{(0)}}
             \times \frac{\nabla \brhog^{(1)}}{\brhog^{(0)}}
    \right]
\nonumber \\
& &
- \frac{1}{4 (\bar{\nu}^{(0)})^2}
  \left[ \nabla \left( \frac{\brhog^{(1)}}{\brhog^{(0)}}
                       + \frac{2 \bar{\nu}^{(1)}}{\bar{\nu}^{(0)}}
                \right)
         \times \frac{\partial_t \nabla \rhog^{(1)}}{\rhog^{(0)}}
         +
         \nabla \left( \frac{\partial_t \brhog^{(1)}}{\brhog^{(0)}}
                       + \frac{\partial_t \bar{\nu}^{(1)}}{\bar{\nu}^{(0)}}
                \right)
         \times \frac{\nabla \brhog^{(1)}}{\brhog^{(0)}}
  \right]
\nonumber \\
&=& - \frac{1}{16 (\bar{\nu}^{(0)})^2}
      \frac{4 + 3 \bar{R}^{(0)}}{1 + \bar{R}^{(0)}}
      \frac{\nabla \brhog^{(1)}}{\brhog^{(0)}} \times
      \frac{\partial_t \nabla \brhog^{(1)}}{\brhog^{(0)}}
    + \frac{1}{16 (\bar{\nu}^{(0)})^2} \frac{\bar{R}^{(0)}}{1 + \bar{R}^{(0)}}
      \frac{\nabla \brhog^{(1)}}{\brhog^{(0)}} \times
      \int dt ~ \frac{\nabla (\nabla^2 \brhog^{(1)})}{\brhog^{(0)}},
\end{eqnarray}
where we used Eq. (\ref{nu0}) and the following relation derived from
Eq. (\ref{TCA1density}),
\begin{equation}
\frac{\nu^{(I,1)}}{\bar{\nu}^{(0)}}
= \frac{4 + 3 \bar{R}^{(0)}}{4(1 + \bar{R}^{(0)})} \Dg^{(I,1)}
  - \frac{\bar{R}^{(0)}}{4(1 + \bar{R}^{(0)})} \frac{1}{\bar{\nu}^{(0)}}
    \int dt ~ \frac{\nabla^2 \brhog^{(1)}}{\brhog^{(0)}}.
\end{equation}
Adding these two contributions, the rotation of the Thomson term can
be written as,
\begin{equation}
\nabla \times \vec{C}^{(2)}
= \frac{1}{16 (\bar{\nu}^{(0)})^2} \frac{1-\beta^3}{1+\beta}
  \frac{\sigma_T \brhog^{(0)}}{e} \frac{\bar{R}^{(0)}}{1 + \bar{R}^{(0)}}
  \frac{\nabla \brhog^{(1)}}{\brhog^{(0)}} \times
  \left[ \frac{\partial_t \nabla \brhog^{(1)}}{\brhog^{(0)}}
         + \int dt ~ \frac{\nabla (\nabla^2 \brhog^{(1)})}{\brhog^{(0)}}
  \right].
\end{equation}

Below we show the leading order quantity for electromagnetic
quantities.  All quantities except the magnetic field need just
$\dvg^{(I,1)}$. On the other hand, as we saw above, the magnetic field
vanishes at the first order in tight coupling approximation and
the second order terms, i.e., 
$\dvg^{(II,1)}$ and $\dvg^{(II,2)}$ are necessary to have a nonzero slip
term and vorticity difference in Eq. (\ref{slip-vorticity}),
respectively.
\begin{eqnarray}
\rho^{(1)}
&=& - \frac{1}{4} \frac{1-\beta^3}{1+\beta^2} \frac{1}{1+\Rbz} \frac{\mp}{e}
      \frac{\nabla^2 \brhog^{(1)}}{\brhog^{(0)}}, ~~~~~
\dn^{(1)}
= - \frac{1}{4} \frac{1-\beta^3}{1+\beta^2} \frac{1}{1+\Rbz} \frac{\mp}{e^2}
    \frac{\nabla^2 \brhog^{(1)}}{\brhog^{(0)}}, \\
\vec{j}^{(1)}
&=& \frac{1}{4} \frac{1-\beta^3}{1+\beta^2} \frac{1}{1+\Rbz} \frac{\mp}{e}
    \frac{\partial_t \nabla \brhog^{(1)}}{\brhog^{(0)}}, ~~~~~
\dv^{(1)}
= \frac{1}{4} \frac{1-\beta^3}{1+\beta^2} \frac{1}{1+\Rbz} \frac{\mp}{e^2 \nb}
  \frac{\partial_t \nabla \brhog^{(1)}}{\brhog^{(0)}}, \\
\vec{E}^{(1)}
&=& - \frac{1}{4} \frac{1-\beta^3}{1+\beta^2} \frac{1}{1+\Rbz} \frac{\mp}{e}
      \frac{\nabla \brhog^{(1)}}{\brhog^{(0)}}, \\
\vec{B}^{(2)}
&=& - \frac{1}{16} \frac{(1+\beta)(1-\beta^3)}{(1+\beta^2)^2}
      \frac{\bar{R}^{(0)}}{(1 + \bar{R}^{(0)})^3}
      \frac{\mp^2}{e \sigma_T \brhog^{(0)}}
      \int dt ~ \frac{\nabla \brhog^{(1)}}{\brhog^{(0)}} \times
                \left[ \frac{\partial_t \nabla \brhog^{(1)}}{\brhog^{(0)}}
                       + \int dt ~ \frac{\nabla(\nabla^2 \brhog^{(1)})}{\brhog^{(0)}}
                \right].
\label{magnetic}
\end{eqnarray}
We see that all quantities are expressed by some background quantities
and a single perturbed quantity, $\brhog^{(1)}$. In
Eq. (\ref{magnetic}), the first term in the bracket is contributed
from both the slip term and vorticity difference, while the second
term is contributed only from the vorticity difference.

\section{Discussion and Summary \label{section:Discussion}}

In this paper, we made a physical interpretation of the results of
previous studies on magnetic field generation from density
fluctuations in the pre-recombination era. This was done by solving
Maxwell equations, Ohm's law and an equation for velocity difference
between photons and baryons. First we expressed electromagnetic
quantities in terms of the Thomson term and studied their
behavior. We saw that timescales for Coulomb scattering are so
short that charge distribution quickly relaxes into its equilibrium
state, which is not charge neutrality but charge separation
which balances with the external force from photons. It was also
shown that magnetic field and electric current do not vanish
even after the source term disappear in contrast to electric
field and charge density. Then the Thomson term was obtained by
the tight coupling approximation up to second order and electromagnetic
quantities were expressed by conventional quantities such as
the density fluctuations of photons. We found that the second order terms
in the tight coupling play essential roles for generation of
the magnetic field.  

Let us give some order-of-magnitude estimation of various quantities obtained
in this paper. The deviation of motion between photons and baryons is
evaluated as,
\begin{eqnarray}
\left| \frac{3}{4} \Dg^{(I,1)} - \Db^{(I,1)} \right|
&\sim& \left| \dvg^{(1)} \right|
\sim \frac{1}{4} \frac{\mp}{\sigma_T \rhog^{(0)}} k \delta_{\gamma}
= \frac{1}{4} k \tau_{\rm T} \delta_{\gamma}
\nonumber \\
&\sim& 3 \times 10^{-10} \left( \frac{k}{k_{\rm Silk}} \right)
       \left( \frac{1+z}{10^{5}} \right)^{-3/2}
       \left( \frac{\deltag}{10^{-5}} \right).
\label{GBvalue}
\end{eqnarray}
This is exactly what is expected from the tight coupling approximation.
On the other hand, the magnitudes of electromagnetic quantities are,
\begin{eqnarray}
\left| \frac{\rho^{(1)}}{e \nb^{(0)}} \right|
&\sim& \left| \dv^{(1)} \right|
\sim \frac{\mp}{4 e^{2} \bnb^{(0)}} k^2 \deltag
= \frac{1}{4 \beta} \frac{k^2}{\omega_p^2} \deltag
\nonumber \\
&\sim& 3 \times 10^{-34} \left( \frac{k}{k_{\rm Silk}} \right)^2
       \left( \frac{1+z}{10^{5}} \right)^2
       \left( \frac{\deltag}{10^{-5}} \right).
\label{EMvalue}
\end{eqnarray}
This is much smaller than expected from the tight coupling approximation
for Coulomb scattering. The reason for this can be seen in section
\ref{subsection:Maxwell-1}, where we used two types of approximation.
The first is the conventional tight coupling approximation for
Coulomb scattering, Eq. (\ref{suppress}), and the second is
Eq. (\ref{suppress2}). Thus electromagnetic quantities are suppressed
by two factors,
\begin{equation}
\frac{\me k}{e^2 \nb \eta_{\rm eff}} \times k \eta_{\rm eff}
= \frac{k^2}{\omega_p^2},
\end{equation}
which is the suppression factor seen in Eq. (\ref{EMvalue}).
The second suppression factor can be attributed to the existence
of coupling due to the electric field. Equations (\ref{GBvalue})
and (\ref{EMvalue}) justify our strategy on the tight coupling
approximation that we solved up to second and first order for
Thomson and Coulomb scattering, respectively, dropping
the nonlinear terms in $\dn$ and $\dv$ and keeping the nonlinear
terms in $\dvg$, $\Dg$ and $Db$ up to second order.

It is instructive to know how equations of motion for photons (\ref{gammaEOM}),
protons (\ref{pEOM}) and electrons (\ref{eEOM}) are balanced.
each term of the equations at the leading order. \\
photons:
\begin{equation}
{\rm (acceleration):(pressure):(Thomson(with~protons)):(Thomson(with~electrons))}
\sim 1:1:\beta^2 R:R,
\end{equation}
protons:
\begin{equation}
{\rm (acceleration):(electric~field):(Thomson):(Coulomb)}
\sim 1:1:\beta^2:k \eta,
\end{equation}
electrons:
\begin{equation}
{\rm (acceleration):(electric~field):(Thomson):(Coulomb)}
\sim \beta:1:1:k \eta.
\end{equation}
It can be seen that Coulomb scattering is not important for cosmological scales
with $k \eta \ll 1$, which corresponds to the fact that magnetic diffusion
is absent at these scales.

In this paper, while we ignored the cosmological expansion and the
anisotropic stress of photons, which makes this treatment rather not
quantitative but qualitative, we could develop a comprehensive
treatment of photon, electron and proton fluids, and successfully
understand how the deviation between these fluids are determined.
We found out deviation between photons and charged particles is much larger
than that of protons and electrons. Our analytic treatment is particularly
important on scales much smaller than the cosmological horizon where we can
safely ignore the cosmic expansion and the numerical analysis is rather
difficult to carry out. To make a precise prediction of the spectrum of
the magnetic field, however, we need to evaluate the contribution from
vorticity difference together with the anisotropic stress of photons,
which we will present in future.

\acknowledgements

KT and KI are supported by a Grant-in-Aid for the Japan Society for
the Promotion of Science Fellows and are research fellows of the Japan
Society for the Promotion of Science. NS is supported by a Grant-in-Aid
for Scientific Research from the Japanese Ministry of Education (No. 17540276).
KT would like to thank R. Kulsrud, R. Maartens, Y. Ohira, M. Sasaki,
T. Shiromizu and A. Taruya for helpful suggestions and useful discussions.

\end{document}